\documentclass[prb,twocolumn,showpacs,preprintnumbers,amsmath,amssymb]{revtex4}
\usepackage{mathrsfs}
\usepackage{graphicx}
\begin{document}

\title{Enhancement of ferromagnetism by p-wave Cooper pairing in superconducting ferromagnets}

\author{Xiaoling Jian, Jingchuan Zhang, and Qiang Gu}\email{qgu@sas.ustb.edu.cn}
\affiliation{Department of Physics, University of Science and
Technology Beijing, Beijing 100083, China}
\author{Richard A. Klemm}
\affiliation{Department of Physics, University of Central Florida,
Orlando, Florida 32816, USA}

\date{\today}

\begin{abstract}

In superconducting ferromagnets for which the Curie temperature
$T_{m}$ exceeds the superconducting transition temperature $T_{c}$,
it was  suggested that ferromagnetic spin fluctuations could
lead to superconductivity with $p$-wave spin
triplet Cooper pairing. Using the Stoner model of itinerant ferromagnetism,
we study the feedback effect of the $p$-wave superconductivity on
the ferromagnetism. Below $T_{c}$, the ferromagnetism
is enhanced by the  $p$-wave superconductivity. At zero
temperature, the critical Stoner value for itinerant ferromagnetism
is reduced by the strength of the $p$-wave pairing potential, and
the magnetization increases correspondingly. More important, our results suggest that once Stoner ferromagnetism is established, $T_m$ is unlikely to ever be below $T_c$.
 For strong and weak ferromagnetism,  three and two
peaks in the temperature dependence of the specific heat are
respectively predicted, the upper peak in the latter case corresponding to a first-order transition.
\end{abstract}


\maketitle

Due to the strong interplay between conventional superconducting
(SC) and ferromagnetic (FM) states, the exploration of their
possible coexistence in the same crystal might have seemed
fruitless, but has nevertheless attracted a great deal of interest
recently. This possible coexistence was first proposed by Ginzburg
more than 50 years ago\cite{Ginzburg1957}. Several years later,
Larkin and Ovchinnikov\cite{Larkin1964} and Fulde and
Ferrell\cite{Fulde1964} independently developed a microscopic theory
of this coexistence in the presence of a strong magnetic field,
based upon a spatially inhomogeneous SC order parameter, presently
referred to as the FFLO state. Meanwhile, Berk and Schrieffer
suggested that conventional $s$-wave superconductivity in the
paramagnetic phase above the Curie temperature $T_{m}$ is suppressed
by critical ferromagnetic fluctuations near to
$T_{m}$\cite{Berk1966}. However, more recent calculations showed
that conventional $s$-wave superconductivity can form in the weakly
FM regime close to a quantum phase transition\cite{Blagoev1999}. In
addition, Fay and Appel predicted that $p$-wave superconductivity
could arise in itinerant ferromagnets\cite{Fay1980}. Their
pioneering work indicated that longitudinal ferromagnetic spin
fluctuations could result in a $p$-wave ``equal-spin-pairing" SC
state within the FM phase.

Experimentally, a major development occurred with the observation by
Saxena {\it et al.} that ${\rm UGe_{2}}$, nominally an itinerant FM
compound, undergoes an SC transition at low $T_{c}$ values under
high pressure\cite{Saxena2000}. An SC state was also found in other
itinerant ferromagnets such as ${\rm ZrZn_{2}}$ and ${\rm
URhGe}$\cite{Pfleiderer2001,Aoki2001}. In each case, the regime of
the SC phase appears completely within that of the FM phase,
suggesting a cooperative effect between the SC and FM states.

These experimental achievements have stimulated renewed theoretical
interest in the subject. Recently, a large effort has been devoted
to the understanding of the underlying physics of the coexisting SC
and FM states, with a focus upon the SC pairing mechanism and the
orbital symmetry of the SC order parameter. Although earlier works
by Suhl and Abrikosov suggested that an $s$-wave pairing interaction
between conduction electrons could be mediated by
ferromagnetically-ordered localized spins, such as by
impurities\cite{Suhl2001,Abrikosov2001}, recent studies of these SC
ferromagnets\cite{Saxena2000,Pfleiderer2001,Aoki2001} have assumed
that the itinerant electrons involved in both the FM and SC states
are within the same
band\cite{Karchev2001,Shen2003,Cuoco2003,Kirkpatrick2001,Machida2001,Nevidomskyy2005}.
Some of these studies assumed conventional $s$-wave pairing. For
example, Karchev {\it et al.} studied an itinerant electron model
in which the same electrons are responsible for both the FM and SC
states\cite{Karchev2001}. In that study, the Cooper pairs were
assumed to be in a spin-singlet state, and the ferromagnetism was
described within the Stoner model. However, the resulting SC
ferromagnetic state was shown to be energetically unfavorable when
compared to the conventional, nonmagnetic SC state\cite{Shen2003}. A
possible exception to this incompatibility could occur if the
magnetic instability were to arise from a dynamic spin exchange
interaction, as discussed by Cuoco {\it et al.}\cite{Cuoco2003}. On
the other hand, a number of other workers avoided the likely
incompatibility of the SC and FM states by assuming a spin-triplet
SC order parameter with $p$-wave orbital symmetry, for
simplicity\cite{Kirkpatrick2001,Machida2001,Nevidomskyy2005}.
Kirkpatrick {\it et al.} indicated that a $p$-wave SC state
meditated by ferromagnetic spin fluctuations is more likely to
coexist within the Heisenberg FM phase regime than within the
paramagnetic phase regime\cite{Kirkpatrick2001}. Machida and Ohmi
studied the properties of a $p$-wave SC ferromagnet
phenomenologically\cite{Machida2001}. More recently, a microscopic
model of the coexistence of a nonunitary spin-triplet SC state with
a weakly itinerant FM state was developed by
Nevidomskyy\cite{Nevidomskyy2005}. The present nature of the SC
coexistent with the FM state in these ferromagnetic superconductors
is still somewhat controversial, although increasingly, additional
experiments on the U-based materials have provided increasing
support for a spin-triplet state rather than a spin-singlet
one\cite{Huxley2001,Hardy2005,Harada2007,Huy2007}.

Most theoretical studies have focused primarily on the effect of the
established ferromagnetism upon the nature of the coexistent
superconductivity, as summarized above. However, to fully understand
the interplay between the SC and FM states when they coexist, one
should also study the feedback effect of the superconductivity upon
the ferromagnetism itself, as has been done in only one study to
date\cite{Nevidomskyy2005}.

Here we study explicitly the effects of the $p$-wave pairing on the
FM ordering, using the Stoner model of itinerant ferromagnetism as
the starting point. We calculate the critical Stoner parameter
$U_{c}$, the magnetization $m$, and the two parallel-spin $p$-wave
gap function magnitudes, $\Delta_{\pm}$, respectively, as functions
of the pair-interaction strength $V$. We also discuss
finite-temperature properties, including the $T$-dependencies of
these order parameters and the specific heat $C(T)$.

We take the Hamiltonian  for the ferromagnetic superconductor to have the
form
\begin{eqnarray}
H_{FM+SC}&=& \sum_{\mathbf{k},\sigma}(\epsilon_{\mathbf{k}} - \mu
   -\sigma M) c_{\mathbf{k}\sigma}^{\dag}c_{\mathbf{k}\sigma}\nonumber\\
  &+&\frac{1}{2\mathcal{V}}\sum_{\substack{\mathbf{k},\mathbf{k}^{\prime}\\
   \sigma,\sigma^{\prime}}}V_{\rm SC}(\mathbf{k},\mathbf{k}^{\prime})
   c_{\mathbf{k},\sigma}^{\dag}c_{\mathbf{-k},\sigma^{\prime}}^{\dag}
   c_{\mathbf{-k}^{\prime}\sigma^{\prime}}c_{\mathbf{k}^{\prime}\sigma},
\end{eqnarray}
where $\sigma=\pm$ represent the single-particle spin states, and
the single-quasiparticle part of $H$ comprises the Stoner model for
itinerant electrons, where $\epsilon_{\mathbf{k}}$ is the
non-magnetic part of the quasiparticle dispersion, $\mu$ is the
chemical potential and $M=U(\langle n_{+}\rangle-\langle
n_{-}\rangle)/2$ is the magnetic molecular-field with $U$ the Stoner
exchange interaction, and $\mathcal{V}$ is the sample volume. The
pairing potential is taken to have the $p$-wave form\cite{Mahan},
$V_{\rm SC}(\mathbf{k},\mathbf{k}^{\prime}) = -
V\hat{\mathbf{k}}\cdot\hat{\mathbf{k}}^{\prime}$. In  weak coupling
theory, $V$ is non-zero and assumed to be  constant only within the
narrow energy region $|\epsilon-\epsilon_F|\le \omega_{c}$ near to
the Fermi energy $\epsilon_F$,
 where $\omega_{c}$ is the
energy cut-off.

Because of the pair-breaking effects of the strong exchange field in
ferromagnets, we assume that only parallel-spin Cooper pairs can
survive. Thus we set the $p$-wave antiparallel-spin gap function $\Delta_{0}=0$ and
retain the two gap functions with parallel-spin states $m_S=\pm1$, $\Delta_{\pm1}$.
The SC order parameter is assumed to have the following $p$-wave
symmetry\cite{Mahan}, $\Delta_{\pm1}({\bf k}) =
(\hat{\mathbf{k}}_{x}+i\hat{\mathbf{k}}_{y}) \Delta_{\pm}.$

The Hamiltonian is treated via the Green function method within the
mean-field theory framework. In addition to the normal Green
function $\mathscr{G}_{\sigma}(\mathbf{k},\tau-\tau^{\prime}) =
-\langle
T_{\tau}c_{\mathbf{k}\sigma}(\tau)c^{\dag}_{\mathbf{k}\sigma}(\tau^{\prime})
\rangle$, the anomalous Green function
describing the pairing of electrons should be introduced,
$\mathscr{F}_{\sigma}(\mathbf{k},\tau-\tau^{\prime}) = \langle
T_{\tau}c_{\mathbf{k}\sigma}(\tau)c_{\mathbf{-k}\sigma}(\tau^{\prime})
\rangle$. Using the standard equation of motion approach, the Green
functions are derived to be
\begin{equation}\label{GF}
\begin{split}
\mathscr{G}_{\pm}(\mathbf{k},ip_{n})&= \frac{-(ip_{n} +
    \epsilon_{\mathbf{k}}\mp M)}{p_{n}^{2} +
    (\epsilon_{\mathbf{k}}\mp M)^{2}+|\Delta_{\pm1}({\bf k})|^{2}},\\
\mathscr{F}_{\pm}(\mathbf{k},ip_{n})&=\frac{\Delta_{\pm1}}
    {p_{n}^{2}+(\epsilon_{\mathbf{k}}\mp M)^{2}+|\Delta_{\pm1}({\bf k})|^{2}},
\end{split}
\end{equation}
where the $p_n$ are the Matsubara frequencies, and the FM and SC order parameters  are respectively defined as
\begin{equation}\label{OP}
\begin{split}
M&=\frac{U}{2\mathcal{V}}\sum_{\mathbf{k}}(\langle
   n_{\mathbf{k}+}\rangle-\langle n_{\mathbf{k}-}\rangle),\\
\Delta_{\pm1}(\mathbf{k})&= -{\frac{1}{\mathcal{V}}}
   \sum_{\mathbf{k}^{\prime}}V_{\rm SC}(\mathbf{k},\mathbf{k}^{\prime})
   \mathscr{F}_{\pm}(\mathbf{k}^{\prime},\tau=0).
\end{split}
\end{equation}
All of the order parameters can be calculated using the above Green
functions. They are found to satisfy
\begin{equation}\label{mag1}
M = \frac{U}{2\mathcal{V}}\sum_{\mathbf{k}}
    \left\{\frac{\epsilon_{\mathbf{k}}^{\uparrow}[1 - 2f(E_{-})]}{2E_{-}(\mathbf{k}) } -
    \frac{\epsilon_{\mathbf{k}}^{\downarrow}[1-2f(E_{+})]}{2E_{+}(\mathbf{k})}\right\},
\end{equation}
\begin{equation}\label{sup}
\Delta_{\pm1}(\mathbf{k}) = \frac{-1}{\mathcal{V}}
   \sum_{\mathbf{k}^{\prime}} V_{\rm SC}(\mathbf{k},\mathbf{k}^{\prime})\frac{1 -
   2f[E_{\pm}({\mathbf{k}}^{\prime})]}{2 E_{\pm}({\mathbf{k}}^{\prime})}
   \Delta_{\pm1}({\mathbf{k}}^{\prime}),\\
\end{equation}
where $\epsilon_{\mathbf{k}}^{\uparrow,\downarrow} =
\epsilon_{\mathbf{k}} - \mu \pm M$, $E_{\pm}(\mathbf{k})=
\sqrt{(\epsilon_{\mathbf{k}}^{\downarrow,\uparrow})^{2} +
|\Delta_{\pm1}(\mathbf{k})|^{2}}$, and $f(E)$ is the Fermi function.
The chemical potential $\mu$ is determined from the equation for the
number of electrons per unit volume, or particle density,
\begin{equation}\label{n1}
n = \frac{1}{\mathcal{V}}\sum_{\mathbf{k}}\left\{1 -
  \frac{\epsilon_{\mathbf{k}}^{\uparrow}[1-2f(E_{-})]}{2E_{-}(\mathbf{k})} -
  \frac{\epsilon_{\mathbf{k}}^{\downarrow}[1-2f(E_{+})]}{2E_{+}(\mathbf{k})}\right\},
\end{equation}
which is
equal to unity at half filling.

Equations (\ref{mag1}), (\ref{sup}) and (\ref{n1}) with $n=1$ comprise the
self-consistent equations for the ferromagnetic superconducting
system. We solve the equations for the simple case of a spherical
Fermi surface at half filling. It is convenient to
solve these equations  by converting the summations over $\bf{k}$-space to
continuum integrals over energy,
\begin{eqnarray}
\overline{M} &=& \frac{\overline{U}}{32\pi^{2}}\int^{\infty}_{0}d\overline{\varepsilon}
   \int^{\pi}_{0} d\theta\sin\theta\sqrt{\overline{\varepsilon}} \nonumber \\
  && \quad\;\;\; \times\left\{
   \frac{\overline{\varepsilon}^{\uparrow}\tanh[\frac{\overline{E}_{-}}{2\overline{T}}]}
   {\overline{E}_{-}} - \frac{\overline{\varepsilon}^{\downarrow}
   \tanh[\frac{\overline{E}_{+}}{2\overline{T}}]}{\overline{E}_{+}}\right\}, \label{mag2} \\
1 &=& \frac{\overline{V}}{32\pi^{2}}\int^{\overline{\epsilon}_{F_{\pm}} +
   \overline{\omega}_{c}}_{\overline{\epsilon}_{F_{\pm}}-\overline{\omega}_{c}}
   d\overline{\varepsilon}\int^{\pi}_{0}d\theta \nonumber \\
  && \qquad\qquad\quad\; \times \left\{\frac{\sqrt{\overline{\varepsilon}}\cdot\sin^{3}\theta}
   {\overline{E}_{\pm}} \tanh[\frac{\overline{E}_{\pm}}{2\overline{T}}]\right\}, \label{supdm2} \\
n &=&\frac{1}{16\pi^{2}}\int^{\infty}_{0}d\overline{\varepsilon}\int^{\pi}_{0}
   d\theta\sin\theta\sqrt{\overline{\varepsilon}} \nonumber \\
  && \times\left\{2-\frac{\overline{\varepsilon}^{\uparrow}
   \tanh[\frac{\overline{E}_{-}}{2\overline{T}}]}{\overline{E}_{-}}
   -\frac{\overline{\varepsilon}^{\downarrow}\tanh[\frac{\overline{E}_{+}}{2\overline{T}}]}
   {\overline{E}_{+}}\right\}, \label{n2}
\end{eqnarray}
where $\overline{\epsilon}_{F_{\pm}}=\overline{\mu}\pm
\overline{M}$,
$\overline{\varepsilon}^{\downarrow,\uparrow}=\overline{\varepsilon}
- \overline{\epsilon}_{F_{\pm}}$, and
$\overline{E}_{\pm}=\sqrt{(\overline{\varepsilon}^{\downarrow,\uparrow})^{2}
+ \sin^{2}\theta|\overline{\Delta}_{\pm}|^{2}}$. In the above
equations, the unit of energy is rescaled by the factor
$\frac{\hbar^{2}n^{2/3}}{2m^{\ast}}$.  The dimensionless
interactions $\overline{U}$ and $\overline{V}$ are thus defined by
$\overline{U} = U (\frac{\hbar^{2}n^{2/3}}{2m^{\ast}})^{-1}$ and
$\overline{V} = V (\frac{\hbar^{2}n^{2/3}}{2m^{\ast}})^{-1}$, and
the dimensionless energies $\overline{\epsilon}_{F_{\pm}}$,
$\overline{\varepsilon}$, $\overline{\omega}_{c}$,
$\overline{E}_{\pm}$, $\overline{\Delta}_{\pm}$, and
$\overline{\mu}$ are defined analogously. The dimensionless
temperature is defined by $\overline{T} =
k_{B}T(\frac{\hbar^{2}n^{2/3}}{2m^{\ast}})^{-1}$.
 We choose
$\overline{\omega}_{c}=0.01\overline{\epsilon}_{F}$, where $\overline{\epsilon}_{F}$ is the dimensionless
Fermi energy at $\overline{M}=\overline{T}=0$.

By solving the  equations self-consistently, we can investigate the
interplay between the magnetism and the superconductivity in the
coexisting state. This issue was discussed  previously based
on a similar framework, with the emphasis placed on the effects on the SC
pairing due to the critical spin fluctuations in FM
compounds\cite{Nevidomskyy2005}. The present work focuses on the
reciprocal action, i.e., the influence of the SC on the FM.

\begin {figure}[t]
\includegraphics[width=0.4\textwidth,keepaspectratio=true]{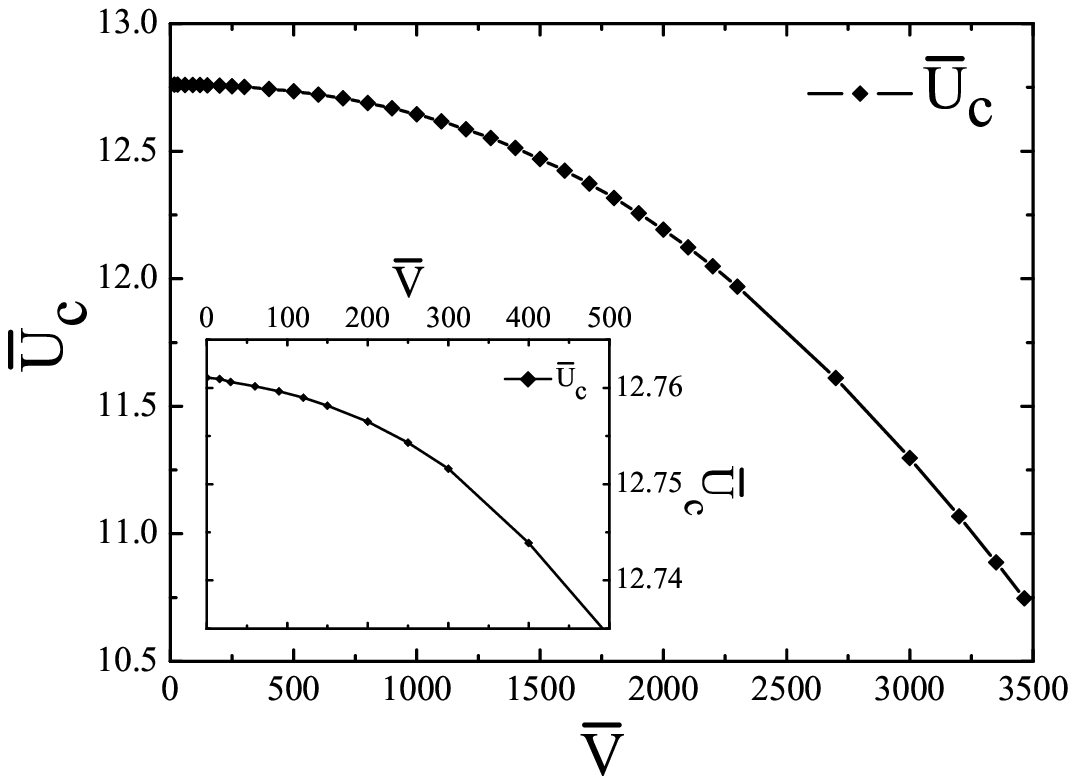}
\caption{The Stoner point $\overline{U}_{c}(\overline{V})$ as a
function of the $p$-wave interaction strength $\overline{V}$ at
$\overline{T}=0$. Inset: Enlargement of the region
$0\le\overline{V}\le500$.} \label{VUc}
\end{figure}

According to Stoner theory, a Fermi gas can exhibit ferromagnetism
only when the effective FM exchange is larger than  the critical
Stoner point. For a system described by  Eq. (1), $U$ represents the
effective exchange interaction. In the absence of the $p$-wave SC
interaction, $\overline{V}=0$, the dimensionless Stoner point
$\overline{U}_{c}(0)\approx 12.76104$. For $\overline{V}\ne0$, we
calculate $\overline{U}_{c}(\overline{V})$. As shown in Fig.
\ref{VUc}, the $\overline{T}=0$ Stoner point
$\overline{U}_{c}(\overline{V})$ decreases as $\overline{V}$
increases, which implies that the $p$-wave Cooper pairing reduces
the barrier to the onset of the magnetization of the Fermi gas. We
note that $\overline{V}$ might be very small in a real system, so
the enhancement effect of the superconductivity on the
ferromagnetism may be very weak. The inset of Fig. \ref{VUc} shows
the details of $\overline{U}_{c}(\overline{V})$ in the region of
small $\overline{V}$, where the decreasing tendency of
$\overline{U}_{c}(\overline{V})$ with increasing $\overline{V}$
still can be seen clearly.

\begin {figure}[t]
\includegraphics[width=0.4\textwidth,keepaspectratio=true]{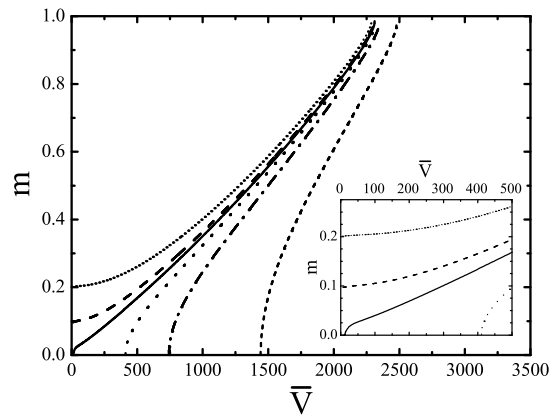}
\caption{Plots of the electronic magnetization $m\equiv\langle n_{+}
\rangle-\langle n_{-}\rangle$ as a function of the $p$-wave
interaction strength $\overline{V}$ at $\overline{T}=0$ for fixed
values of $\overline{U}$. From larger to smaller $m$ at fixed
$\overline{V}$, $\overline{U}=12.8$ (short dotted), 12.77 (dashed),
12.761 (solid), 12.743 (dotted), 12.7 (dash-dotted) and 12.495
(short dashed). Inset: Enlargement of the region
$0\le\overline{V}\le500$.} \label{Vm}
\end{figure}

To further demonstrate the influence of the SC on the FM, we discuss
the magnetization $m\equiv\langle n_{+} \rangle-\langle
n_{-}\rangle$ as a function of $\overline{V}$ at $\overline{T}=0$.
Here we use $m = 2\overline{M}/\overline{U}$ instead of
$\overline{M}$ to eliminate the dependence of $\overline{U}_c$ upon
$\overline{V}$. As shown in Fig. \ref{Vm}, $m(\overline{V})$
increases with $\overline{V}$ for each given value of
$\overline{U}$.  For
 $\overline{U} > \overline{U}_{c}(0)$, $m(0)$ is finite, since the system is spontaneously magnetized, and $m(\overline{V})$ increases  monotonically from $m(0)$, eventually reaching unity at a finite $\overline{V}\le2300$. For
$\overline{U}<\overline{U}_{c}(0)$, however, $m(\overline{V})=0$  for
$\overline{V}<\overline{V}_c(\overline{U})$, and then $m(\overline{V})\ne0$ increases
sharply with  $\overline{V}$ for
$\overline{V}\ge\overline{V}_c(\overline{U})$, eventually reaching
unity at $\overline{V}>2300$. The critical value $\overline{V}_c(\overline{U})$ corresponds
to the reduction in the Stoner point $\overline{U}_c(\overline{V})$ at which the onset of the ferromagnetism is  induced, as pictured in Fig. \ref{VUc}. This is a second way in which the $p$-wave superconductivity can
enhance the ferromagnetism.

\begin {figure}[b]
\includegraphics[width=0.4\textwidth,keepaspectratio=true]{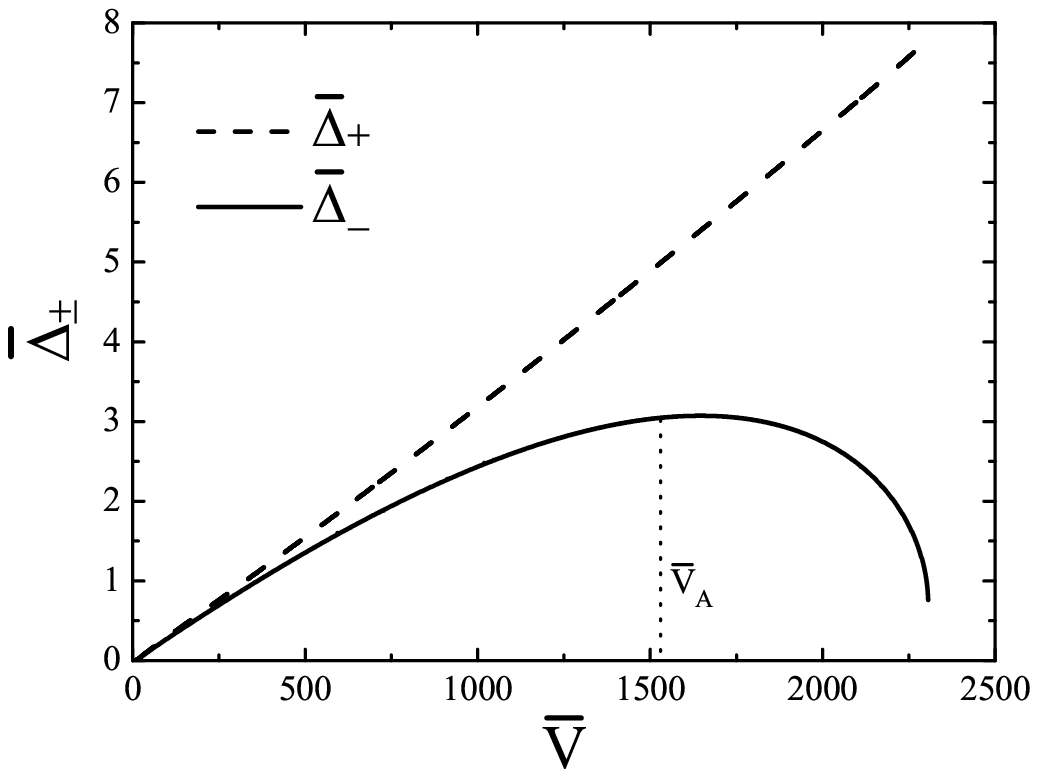}
\caption{Plots of $\overline{\Delta}_{+}$ (dashed) and
$\overline{\Delta}_{-}$ (solid) as functions of $\overline{V}$ at
$\overline{U}=12.77$ and $\overline{T}=0$. $\overline{V}_{A}$ is the
value of $\overline{V}$ at which $\overline{\Delta}_{-}$ has a
maximum, and $\overline{\Delta}_{-}\rightarrow0$ at
$\overline{V}\rightarrow\sim2300$, the point at which
$m\rightarrow1$ in Fig. \ref{Vm}.}
 \label{supDelta}
\end{figure}

A similar effect was found in the ferromagnetic spin-1
Bose gas which exhibits two phase transitions, the FM transition and
Bose-Einstein condensation (BEC). The BEC temperature increases with
FM couplings and, on the other hand, the FM transition is
significantly enhanced due to the onset of the BEC\cite{Gu2004}.
Considering that  triplet Cooper pairs behave somewhat like
spin-1 bosons, a FM superconductor is analogous to a FM Bose
gas.

Figure \ref{supDelta} displays plots of the $p$-wave SC order
parameters, $\overline{\Delta}_{\pm}$  as  functions of $\overline{V}$ at  $\overline{T}=0$ and
$\overline{U}=12.77$, just above the $\overline{V}=0$ Stoner point
$\overline{U}_c(0)$.  Although  with increasing $\overline{V}$,
$\overline{\Delta}_{+}$ rises monotonically,
 $\overline{\Delta}_{-}$
initially rises, reaches a maximum  at $\overline{V}_A$, and then
decreases at an increasing rate until it  vanishes discontinuously when $m(\overline{V})=1$.
For $\overline{U}=12.77$,  $m(\overline{V})>0$
is shown by the
dashed  curve in Fig. \ref{Vm}, so that
$\overline{\Delta}_{+}>\overline{\Delta}_{-}$ for all
$\overline{V}$. Since $m$ also grows with $\overline{V}$, the mean number
of spin-down electrons decreases  with increasing
$\overline{V}$, vanishing when $m\rightarrow1$ at
$\overline{V}\approx2300$, at and beyond which  $\overline{\Delta}_{-}\rightarrow0$.

\begin {figure}[t]
\includegraphics[width=0.4\textwidth,keepaspectratio=true]{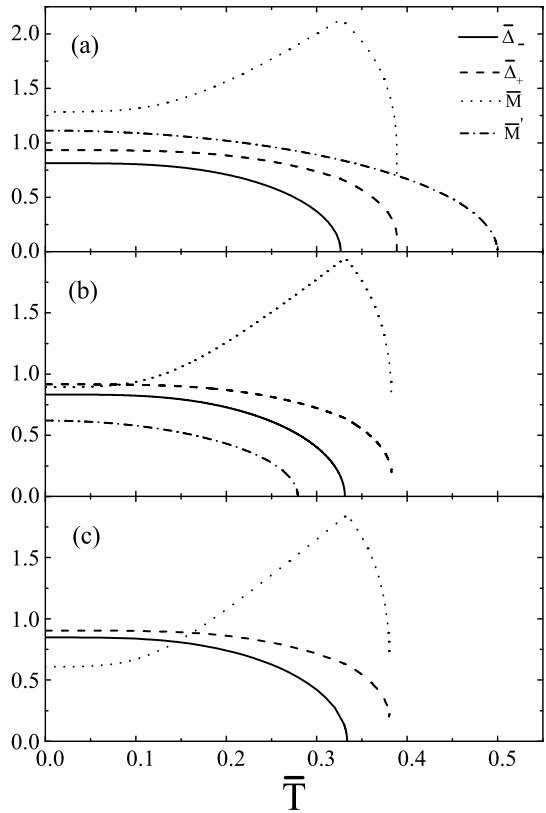}
\caption{Shown are plots of the order parameters $\overline{M}$
(dotted), $\overline{\Delta}_{+}$ (dashed), and
$\overline{\Delta}_{-}$ (solid) as functions of $\overline{T}$ in
the coexistence state for  $\overline{V}=300$.
$\overline{M}^\prime$ (dash-dotted) is the magnetic order parameter
when $\overline{V}=0$. (a) $\overline{U}=12.79> \overline{U}_{c}(0)$
and $\overline{T}_m^\prime>\overline{T}_{c+}$. (b)
$\overline{U}=12.77> \overline{U}_{c}(0)$ but
$0<\overline{T}_m^\prime<\overline{T}_{c+}$. (c)
$\overline{U}=12.76<\overline{U}_{c}(0)$ but
$\overline{U}>\overline{U}_{c}(V)$. The ferromagnetism is induced
due to the $p$-wave pairing ($\overline{M}\ne 0$) even though
$\overline{M}^{\prime}=0$.} \label{Tdp}
\end{figure}

We now discuss the finite temperature properties of the system. We
define $\overline{M}^\prime$ to be the magnetic order parameter when
$\overline{V}=0$, for which $\overline{\Delta}_{\pm}=0$. The
$\overline{T}$ dependencies of the order parameters
$\overline{\Delta}_{\pm}$,
 $\overline{M}$, and $\overline{M}^\prime$
are obtained numerically and shown for $\overline{V}=300$ and three different
$\overline{U}$ cases in Fig. \ref{Tdp}. The order parameters become
non-vanishing below their respective dimensionless transition temperatures
$\overline{T}_{c\pm}$,  $\overline{T}_m$, and $\overline{T}_m^\prime$. In each case, the SC
order parameters $\overline{\Delta}_{\pm}$ increase monotonically
with decreasing $\overline{T}$ below $\overline{T}_{c\pm}$, respectively. In
the FM superconductor, $\overline{T}_{c-} < \overline{T}_{c+}$ and
$\overline{\Delta}_{-}(\overline{T})<\overline{\Delta}_{+}(\overline{T})$,
as shown in Figs. 4(a), 4(b), and 4(c).  In addition,
$\overline{M}^\prime(\overline{T})$ also increases monotonically
with decreasing $\overline{T}$ for the ferromagnet in the absence of
any superconductivity, as depicted in Figs. 4(a) and 4(b) for the
respective cases $\overline{U}>\overline{U}_c(0)$ and
$\overline{T}_m^\prime>\overline{T}_{c+}$ and
$0<\overline{T}_m^\prime<\overline{T}_{c+}$. However, the
$\overline{T}$-dependence of $\overline{M}$  is non-trivial when
$p$-wave superconductivity is present. In the first case pictured in Fig. 4(a),
$\overline{M}(\overline{T})=\overline{M}^\prime(\overline{T})$ for
$\overline{T}_m^\prime>\overline{T}_{c+}$, as in the absence of superconductivity.
However, $\overline{M}(\overline{T})$ exhibits an upward kink at
$\overline{T}_{c+}$ below which
$\overline{\Delta}_{+}\ne0$.  Then, for
$\overline{T}_{c-}<\overline{T}<\overline{T}_{c+}$, $\overline{M}$
increases sharply with decreasing $\overline{T}$, and exhibits a
downward kink at $\overline{T}_{c-}$ below which  $\overline{\Delta}_{-}\ne0$.  Below
$\overline{T}_{c-}$, $\overline{M}(\overline{T})$ then decreases
monotonically with $\overline{T}$. This case was discussed
previously in a similar scenario\cite{Dahal2005}.

The case $\overline{T}_m^\prime<\overline{T}_{c\pm}$ not previously discussed is more interesting.  Two examples of this case with $\overline{V}=300$ are shown in Figs. \ref{Tdp}(b) and \ref{Tdp}(c).  In Fig. 4(b), the
magnetization $\overline{M}^{\prime}$ for $\overline{V}=0$ (and
$\overline{\Delta}_{\pm}=0$)  is so weak
that $0<\overline{T}_m^\prime<\overline{T}_{c-}$, but a non-vanishing $\overline{V}$ enhances the magnetization,
$\overline{M}$, causing the actual dimensionless Curie temperature $\overline{T}_m$ to
equal $\overline{T}_{c+}$, below which both
$\overline{\Delta}_{+}(\overline{T})$ and
$\overline{M}(\overline{T})$ become discontinuously non-vanishing,
signaling a first-order transition. Their behaviors  for
$\overline{T}<\overline{T}_{c+}=\overline{T}_m$ are then
qualitatively similar to that shown in Fig. 4(a), with
$\overline{\Delta}_{-}(\overline{T})\ne0$ for
$\overline{T}<\overline{T}_{c-}$, causing a downward kink in
$\overline{M}(\overline{T})$ at $\overline{T}_{c-}$, below which
$\overline{M}(\overline{T})$ decreases monotonically with
$\overline{T}$.  For the more extreme case when $\overline{U}< \overline{U}_c(0)$ and
$\overline{T}_m^\prime=0$  but $\overline{U}>\overline{U}_c(\overline{V})$
depicted in Fig. 4(c), the behaviors of the three order parameters
are very similar to that shown in Fig. 4(b).

\begin {figure}[t]
\includegraphics[width=0.475\textwidth,keepaspectratio=true]{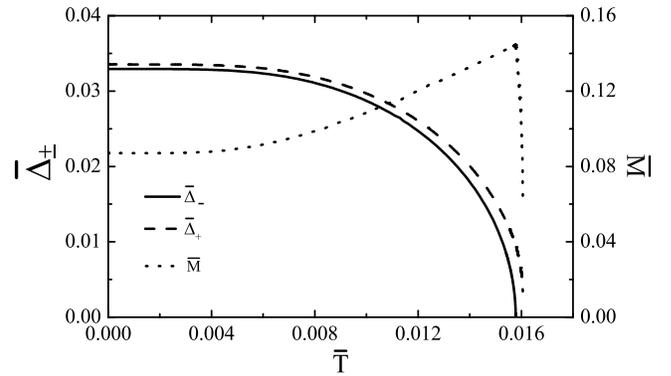}
\caption{Plots of the order parameters $\overline{M}$ (dotted),
$\overline{\Delta}_{+}$ (dashed), and $\overline{\Delta}_{-}$
(solid) as functions of $\overline{T}$ in the coexistence state for
$\overline{V}=20$ and $\overline{U}=12.761<\overline{U}_{c}(0)$. }
\label{Tdp2}
\end{figure}

Considering that $\overline{V}$ is usually small in real systems, a
case with $\overline{V}=20$ is checked, as shown in Fig. 5 where
$\overline{U}$ is taken to be $12.761$, slightly lower than
$\overline{U}_{c}(0)$ but larger than $\overline{U}_{c}(20)\approx
12.7608$. Fig. 5 looks very similar to Fig. 4(c).

Although we did not investigate the limit
$\overline{V}\rightarrow0+$, the examples with $\overline{V}=300$
and $\overline{V}=20$ of the case
$\overline{T}_m^\prime<\overline{T}_{c+}$  pictured in Figs. 4(b),
4(c) and Fig. 5 suggest that in FM superconductors, the actual Curie
temperature $\overline{T}_m$ is unlikely to
 ever be lower than the upper SC transition temperature
$\overline{T}_{c+}$, even if the FM order were extremely weak. In
other words,  these examples argue against the possibility of a FM $\overline{T}$ regime
inside the $p$-wave triplet SC regime, with an actual
$\overline{T}_m<\overline{T}_{c+}$.
Analogously, it was shown that
the ferromagnetic transition never occurs  below the Bose-Einstein
condensation in the FM spin-1 Bose gas\cite{Gu2004}. Moreover, the
present results are to some extent consistent with the observed
phase diagrams of ${\rm UGe_{2}}$\cite{Saxena2000} and ${\rm
ZrZn_{2}}$\cite{Pfleiderer2001}, and with the theoretical discussion
of Walker and Samokhin\cite{Walker2002}, who argued that the
superconductivity only occurs  within the FM region.  In addition, this scenario is consistent with de Haas van Alphen experiments under pressure on UGe$_2$\cite{Terashima}.

However, very recent experiments on UCoGe under pressure were interpreted
as potentially having such a FM regime inside the SC regime near to the FM quantum critical point\cite{deVisser}.  However, the $dc$ resistance and $ac$ susceptibility measurements of $T_m$ and $T_{c+}$  could not determine if there were a FM region inside the SC one for pressures just below their extrapolated quantum critical pressure $p_c$, allowing for a first-order phase transition at the point when $T_m=T_{c+}$, beyond which only a parallel-spin triplet state exists\cite{deVisser}. Further experiments  are encouraged to determine if the FM and SC phase regimes with $0<T_m<T_{c+}$  at fixed pressure actually exist in UCoGe.

\begin {figure}[t]
\includegraphics[width=0.4\textwidth,keepaspectratio=true]{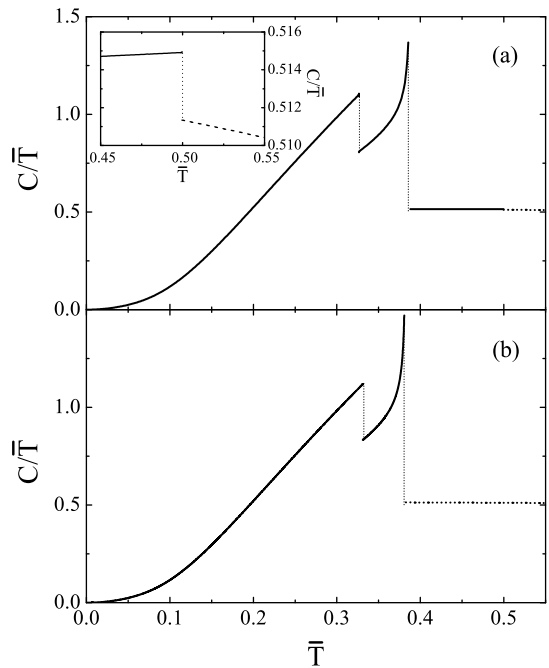}
\caption{Plots of the electronic  specific heat at constant volume
as a function of $\overline{T}$ for (a) a case corresponding to Fig.
\ref{Tdp}(a). The inset shows a transition from ferromagnetic to
paramagnetic phase occurs at the Curie point
$\overline{T}_{m}\approx 0.5$. The dotted curve denotes the specific
heat of the free electron gas; (b) a case corresponding to Fig.
\ref{Tdp}(b). The Curie point $\overline{T}_{m}=\overline{T}_{c+}$, at which the transition is first order.} \label{Cv}
\end{figure}

As suggested by the results for the temperature dependencies of the
order parameters, the FM superconducting system shows multiple phase
transitions, which can be determined experimentally from
measurements of the specific heat. The specific heat at constant
volume for our model can be calculated from
\begin{eqnarray*}
C(\overline{T}) = \overline{T}\frac{\partial S}{\partial
\overline{T}},
\end{eqnarray*}
where the electronic contribution to the entropy  $S$ can be derived from
\begin{eqnarray*}
 S&=&-\sum_{\textbf{k},\sigma=\pm}\{f(\overline{E}_{\sigma})\ln
 f(\overline{E}_{\sigma})\\&&
 +[1-f(\overline{E}_{\sigma})]\ln [1-f(\overline{E}_{\sigma})]\}.
\end{eqnarray*}
The specific heat was calculated previously based on a model of
$s$-wave superconductivity coexisting with
ferromagnetism\cite{J.Jac2003}. For $s$-wave superconductors, there
is only one SC transition temperature $\overline{T}_c$, at which
there is a jump in the specific heat at the second order transition.
However, the case of a $p$-wave superconductor coexisting with
ferromagnetism is more interesting.  In Figs. 6(a) and 6(b), the
results for the specific heat corresponding to the cases pictured in
Figs. 4(a) and 4(b) for the order parameters are shown.  For the
case $\overline{U}>\overline{U}_c$ pictured in Figs. 4(a) and 6(a),
there are three phase transitions at the temperatures
$\overline{T}_{c-}<\overline{T}_{c+}<\overline{T}_m$. In Fig. 6(b),
an example of the case  $\overline{T}_m^\prime<\overline{T}_{c+}$
when $\overline{V}=0$  pictured in Fig. 4(b) is shown. In this case
with $\overline{V}=300$, there is a first-order phase transition at
$\overline{T}_{m}=\overline{T}_{c+}$, and a second-order phase
transition at $\overline{T}_{c-}$.

In conclusion, it is shown that  $p$-wave triplet Cooper pairing can
enhance the ferromagnetism in superconducting ferromagnets. This enhancement is most prominent for the magnetic exchange interaction $U$ very near to the
 Stoner point $U_c(0)$, the critical value for the strength of the exchange
interaction required for  the onset of ferromagnetism in the absence of the $p$-wave pairing interaction $V$.  With finite $V$, $U_c(V)$ is reduced
and the ferromagnetic order parameter increases  in magnitude with
increasing $V$. The temperature dependencies of
the magnetic and parallel-spin superconducting order parameters and
of the specific heat are calculated. The  results show that the Curie temperature is unlikely to ever be lower
than the upper SC transition temperature, in agreement with pressure measurements on UGe$_2$\cite{Terashima}. This feature also may be relevant to recent experiments on UCoGe\cite{deVisser}. The temperature dependence of the specific heat exhibits  two peaks for weak ferromagnetism in the
coexistence state, with a first-order transition at the combined ferromagnetic and upper $p$-wave  SC transition, and  a  lower second-order $p$-wave  SC transition. For strong
ferromagnetism, the specific heat exhibits three second-order transitions.  Our
results support the  possible coexistence of  $p$-wave
superconductivity with a ferromagnetic state.

This work was supported by the Key Project of the Chinese Ministry
of Education (No. 109011), the Fok Yin-Tung Education Foundation,
China (No. 101008), and the NCET program of Chinese Ministry of
Education (NCET-05-0098).  X. J. would like to thank Jihong Qin for
helpful discussions.

\end{document}